\documentclass[a4paper]{article}

\newcommand{\be} {\begin{equation}}
\newcommand{\ee} {\end{equation}}
\newcommand{\bdm} {\begin{displaymath}}
\newcommand{\edm} {\end{displaymath}}
\newcommand{\bc} {\begin{center}}
\newcommand{\ec} {\end{center}}
\newcommand{\beqa} {\begin{eqnarray}}
\newcommand{\eeqa} {\end{eqnarray}}
\newcommand{\nn} {\nonumber}

\newcommand{\bfig} {\begin{figure}}
\newcommand{\efig} {\end{figure}}
\newcommand{\btab} {\begin{tabular}}
\newcommand{\etab} {\end{tabular}}

\font\tenrm=cmr10
\def\half{\textstyle{\frac{1}{2}}}
\def\P{I\!\!P}
\def\R{I\!\!R}
\usepackage{epsfig}
\usepackage{amssymb}
\usepackage{amsmath}
\usepackage[usenames]{color}
\begin{document}
\title{
\rightline{\tenrm MAN/HEP/2014/09}
\rightline{\tenrm DAMTP-2014-43} \bigskip
Central soft production of hadrons in $pp$ collisions}
\author{A Donnachie\\
School of Physics and Astronomy, University of Manchester\\ \\
{P V Landshoff}\\
DAMTP, Cambridge University$^*$\let\thefootnote\relax\footnote{
email addresses: Sandy.Donnachie@hep.manchester.ac.uk, \ 
pvl@damtp.cam.ac.uk}}
\maketitle
\begin{abstract}
The high-energy behaviour of soft scattering observables such as total cross
sections, elastic scattering at small momentum transfer, diffractive
dissociation and central production have been described successfully in the
context of Regge theory, with the same basic structure holding as energies have
increased. For elastic scattering and diffraction dissociation the defining
energies were those of the ISR, the $Sp\bar{p}S$ collider and the Tevatron.
The elastic scattering data from the LHC demonstrate the continuing
applicability of Regge theory. Preliminary data on diffraction dissociation
promise to add to our understanding and now, for the first
time, we can expect to test fully these concepts in central production.
Although the latter is the principal objective of this discussion,
understanding the first two is an essential prerequisite as they define the
formalism and establish parameters.
\end{abstract}
\section{Elastic Scattering}

Regge theory \cite{Collins,DDLN} gives a unified description of soft 
hadronic processes at high energy, in particular elastic scattering at small 
momentum transfer and total cross sections \cite{DL13}. It provides a simple 
quantitative 
description of the combined effect of multiple particle exchanges and has two 
basic components: the exchange of families of
mesons, i.e. $q\bar{q}$ states, and the pomeron that may perhaps be
\cite{Meyer05} the exchange of a family of gluonic $gg$ states. The meson 
families $\R$ lie on trajectories that are to a very good approximation linear,
\be
\alpha_{\R}(t) = \alpha_{\R}(0)+\alpha_{\R}^\prime t
\label{regge trajectory}
\ee
with $\alpha_{\R}(0) \approx 0.5$. They have either $C =-1$ 
($\rho,\omega,\cdots$) or $C=+1$ ($f_2,a_2,\cdots$). As they contribute 
terms to the total cross section that behave as $s^{\alpha_{\R}(0)-1}$,
which is approximately $1/\sqrt{s}$,
by themselves they would
be at variance with observation as all hadronic total cross 
sections increase with $s$ at high energy. The pomeron {$\P$} 
was introduced 
initially to resolve this problem, with a trajectory $\alpha_{\P}$ such that
$\alpha_{\P}(0)=1+\epsilon_{\P}$ with $\epsilon_{\P} > 0$.  As $pp$ and $\bar pp$ total 
ross sections seem to have the same rise with energy at high energy, it has 
$C=+1$. If its trajectory too is approximately linear, then
\be
\alpha_{\P}(t)=1+\epsilon_{\P}+\alpha_{\P}^\prime t.
\label{pomtrajectory}
\ee
The contribution to the elastic amplitude $AB\to AB$ from the exchange of a 
reggeon with trajectory $\alpha(t)$ is
\be
\beta_A(t)\beta_B(t)~\xi^{\pm}(t)~
(s/s_0)^{\alpha(t)}~~~~~~~~~~~~~s_0=1/\alpha'.
\label{elastic}
\ee
Here the {\sl signature factor} $\xi^{\pm}(t)$ is
\be
\xi^{\pm}(t)=1\pm e^{-i\pi\alpha(t)}
\label{signature}
\ee
according to whether the exchange is of $C$-parity $\pm 1$.
The theory does not tell us the coupling $\beta(t)$ of a reggeon to a hadron, 
but data suggest \cite{DDLN} that in each case a good approximation is to take 
it proportional to the hadron's electromagnetic form factor, the Dirac form 
factor in the case of a proton or antiproton. The theory also does not 
identify the fixed scale $s_0$ but the choice $s_0=1/\alpha'$, inspired by 
a clever model introduced long ago by Veneziano \cite{venez} 
is successful for fitting data \cite{DL13}.

In our first fit \cite{sigtot} to total-cross-section data we concluded
that $\alpha_{\P}'=0.25$~GeV$^{-2}$, and
that $\epsilon_{\P}$ should be close to 0.08, though a value closer to
0.096 has been claimed more recently \cite{CKK97}. These are effective values,
representing not just the exchange of a single pomeron, but also
double or more exchanges. This approach works well for describing total cross
sections and elastic scattering at not-too-large $t$, but to extend
the analysis to larger $t$ it is necessary to include explicitly at least
the double exchange $\P\P$. We have concluded \cite{DL13} that this yields
\be
\alpha_{\P}=1.110+0.165t
\label{parameters}
\ee
for the contribution from single-$\P$ exchange.

The term
$\P\P$ is of particular significance as it helps 
explain the 
remarkable dip structure seen in the $p p$ elastic scattering differential 
cross section \cite{Nagy79}. 
The contribution to the amplitude from $\P\P$ exchange has energy dependence 
$s^{\alpha_{\P\P}(t)}$ divided by some function of $\log s$ where, for a linear
pomeron trajectory (\ref{pomtrajectory}),
\be
\alpha_{\P\P}(t)=1+2\epsilon_{\P}+\half\alpha^\prime_{\P}t.
\label{alphac}
\ee
The double-pomeron-exchange contribution is thus flatter in $t$ than single 
exchange, and so its relative importance becomes greater as one goes away from 
$t=0$. It increases more rapidly with energy at $t=0$ than the single exchange,
 and it becomes steeper more slowly, so that the $t$-value beyond which it 
becomes 
important decreases as the energy increases. Interference between the single
and double pomeron exchanges is destructive so the effect is to slow down the
increase of the total cross section with increasing energy. 

It also helps to generate the 
dip and move it to smaller $t$ as the energy increases, in agreement with 
experiment. 
A fundamental property of elastic-scattering amplitudes is that, if they vary
with energy at any value of $t$, they must have complex phase at that $t$.
This is the reason for the signature factor $\xi^{\pm}(t)$ in (\ref{elastic}).
Therefore
modelling the dip structure in detail is complicated as it requires
the simultaneous near-vanishing of both the real and imaginary parts of the
amplitude. The $\P$ and $\P\P$ terms combine to cancel the imaginary part but,
except perhaps at low energies where the meson exchanges can contribute significantly, an additional
term is required to reduce the real part. This can be achieved by introducing
triple-gluon exchange, since this appears \cite{DL79} to dominate the elastic 
amplitude at large values of $t$. In this way, one can obtain an excellent 
description \cite{DL13} of elastic $p p$ and
$\bar{p}p$ scattering from $\sqrt{s}=20$~GeV to 8 TeV.

\section{Diffractive Dissociation}

\begin{figure}[t]
\begin{center}
\epsfxsize=0.4\textwidth\epsfbox[0 0 420 330]{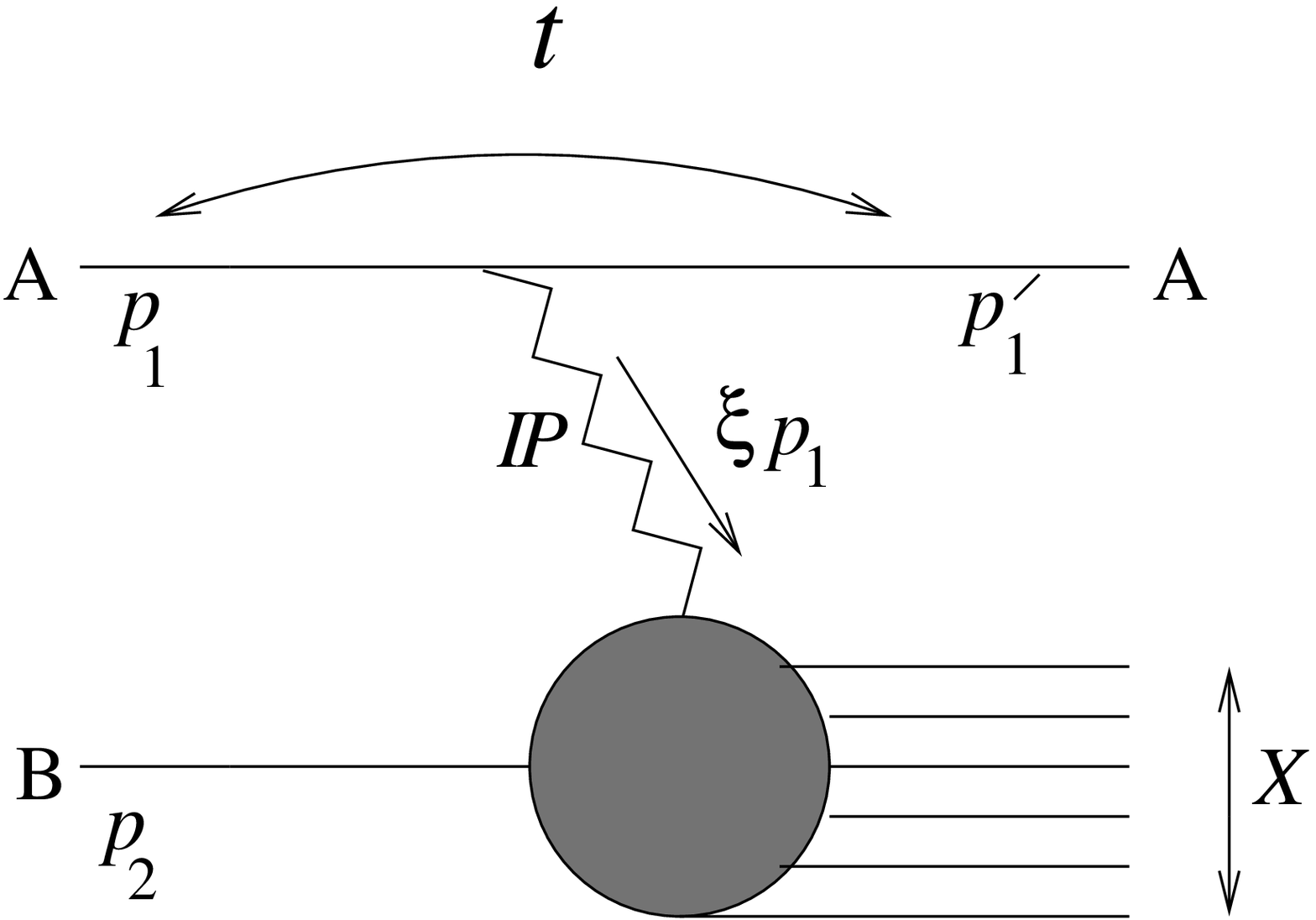}
$\phantom{XXXXXXXXX}$
\epsfxsize35mm\epsfbox[0 -5 280 255]{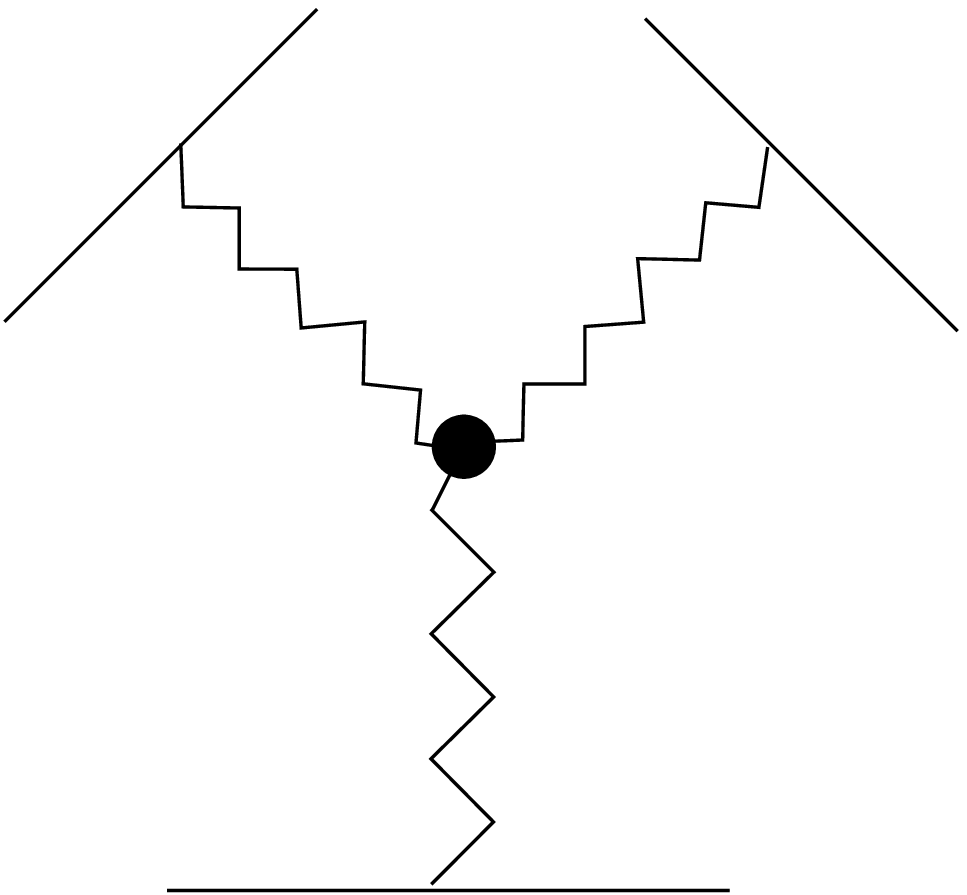}
\vskip 3 truemm
$\hbox{(a)}\phantom{XXXXXXXXXXXXXXXXXXXXXXX}\hbox{(b)}$
\end{center}
\caption{(a) Pomeron exchange in an inelastic diffractive event; (b) the
squared amplitude summed over hadron systems $X$ in the large $M_X^2$ limit}
\label{ddpom}
\end{figure}

We now consider the single-particle inclusive cross section
\be
A(p_1)+B(p_2) \rightarrow A(p_1')+X
\label{dd}
\ee
at high energy and small momentum transfer $t=(p_1'-p_1)^2$ with the mass $M_X$
of the state $X$ large compared with that of the initial hadrons. Then the 
four-momentum $p_1'$ is almost in the same direction as $p_1$ and we may write
$p_1'\sim (1-\xi)p_1$ with $\xi$ small. 
$\xi$ is 
the fractional longitudinal momentum loss of the initial hadron -- not 
to be confused with the signature factor (\ref{signature}).  When $\xi$
is sufficiently small
the process is dominated by pomeron exchange, illustrated in figure 
\ref{ddpom}(a). This may be thought of as a pomeron being ``radiated'' by 
particle $A$ which then interacts with particle $B$ to produce the system $X$.

An inclusive diffractive experiment sums over all possible systems $X$ and so 
measures the total cross section $\sigma^{\P B}(M^2_X,t)$ for a
pomeron of squared mass $t$ scattering on particle $B$. By a generalisation of the optical 
theorem, this is related to the imaginary part of the forward amplitude for 
$\P B$ elastic scattering. The squared centre-of-mass energy corresponding to 
this amplitude is
\be
M_X^2=(\xi p_1+p_2)^2\sim \xi s
\label{defxi}
\ee
and the differential cross section is given by 
\be
\frac{d^2\sigma}{dtd\xi}=D^{\P/A}(t,\xi)~\sigma^{\P B}(M_X^2,t)
\label{diffdissoc}
\ee
where $D^{\P/A}(t,\xi)$ is the pomeron flux and $\sigma ^{\P B}(M_X^2,t)$
is the total $\P B$ cross section. In the notation of (\ref{elastic})
the pomeron flux factor when particle $A$ is a proton or antiproton is
\be
D^{\P/A}(t, \xi)=\frac{[\beta_{p\P}(t)]^2}{4\pi^2}
\xi^{1-2\alpha_{\P}(t)}.
\label{pomflux}
\ee
The pomeron-exchange term (\ref{diffdissoc}) has the 
structure shown in figure \ref{ddpom}(b), in which
three pomerons are coupled together. The lower one carries zero momentum
and the upper two pomerons each carry momentum transfer $t$. In practice,
unless $M_X$ is extremely large,
it is necessary to include meson exchanges $\R$ in $\sigma ^{\P B}(M_X^2,t)$, 
so it has the form
\be
\sigma ^{\P B}(M_X^2,t)=X^{\P B}(t)\,(\alpha_{\P}^\prime M_X^2)^{\alpha_{\P}(0)-1}+Y_+^{\P B}(t)\,(\alpha_{\R}^\prime M_X^2)^{\alpha_{\R}(0)-1}
\label{pomsig}
\ee
where the second term
represents $f_2, a_2$ exchanges. Further, unless $\xi$ is very small, 
we must include terms in which either or both of the upper pomerons are 
replaced with meson exchanges. Thus we need a whole series of terms
\beqa
{\P\P\atop\P}~~~~~~{\P\P\atop f_2}
~~~~~~{f_2\P\atop\P}~~~~~~{\P f_2\atop\P}
~~~~~~{f_2\P\atop f_2}~~~~~~
{\omega\P\atop\omega}~~~~~\dots~.\nn\\ 
\eeqa
with additional contributions also from $\rho$ and $a_2$ exchange, and when $|t|$ 
is of the order of $m_{\pi}^2$ one must also take account of pion exchange.

A term $({ab\atop c})$ contributes to ${d^2\sigma /dt\, d\xi}$
\be
f^A_a(t)f^A_b(t)f^B_c(0)G^{ab}_c(t)\,
e^{i(\phi(\alpha_a(t))-\phi(\alpha_b(t)))}
\xi^{1-\alpha_a(t)-\alpha_b(t)}
\Big ({\alpha^\prime_c M_X^2}\Big )^{\alpha_c(0)-1}.
\label{diffsig}
\ee 
Here, $f^A_a(t)$ and $f^A_b(t)$ are the couplings of the reggeons $a$ and $b$
to the hadron $A$, while $f^B_c(t)$ is that of the reggeon~$c$
to the hadron $B$. $G^{ab}_c(t)$ is the triple-reggeon vertex and 
$\phi(\alpha(t))$ is the phase arising from the signature
factor (\ref{signature}) associated with the trajectory $\alpha(t)$. 
The complex exponential is replaced with 
$2\cos(\phi(\alpha_1(t))-\phi(\alpha_2(t)))$
when we add the term $({ba\atop c})$. Usually, for simplicity, only terms of 
the form 
$({aa\atop c})$ are considered, in which case 
each term contributes
\be
f^A_a(t)f^A_a(t)f^B_c(0)G^{aa}_c(t)\xi^{\alpha_c(0)-2\alpha_a(t)}
(\alpha_c^\prime s)^{\alpha_c(0)-1}
\label{diffsig2}  
\ee
to ${d^2\sigma /dt\, d\xi}$.

Figure \ref{ddpom}b needs adding to it terms in which the single exchanges
are replaced by double and higher exchanges. As we have explained above, it is assumed that this can be
modelled well by using an effective value for $\alpha_{\P}(0)$ less than 1.1
rather than that given in (\ref{parameters}). However, this does not take account of additional exchanges in figure \ref{ddpom}b directly between the intial or final hadrons. Their importance is not known.

The ``total diffractive cross section''
\be 
\sigma^{\rm Diff}(s) = \int_{t_{\rm min}}^{t_{\rm max}}dt 
\int_{\xi_{\rm min}}^{\xi_{\rm max}} d\xi\frac{d^2d\sigma}{dtd\xi}
\label{ddtot}
\ee
is frequently quoted, although it is not a total cross section in the usual 
sense as it is integrated over finite ranges of $t$ and $\xi$. 
Usually the 
single-arm cross section is multiplied by a factor of 2 to include the 
contribution 
\be
A(p_1)+B(p_2) \rightarrow X+B(p_2')
\ee
with the same $t$ and $\xi$. 

\bfig[t]
\bc
\begin{minipage}{80mm}
\epsfxsize80mm
\epsffile{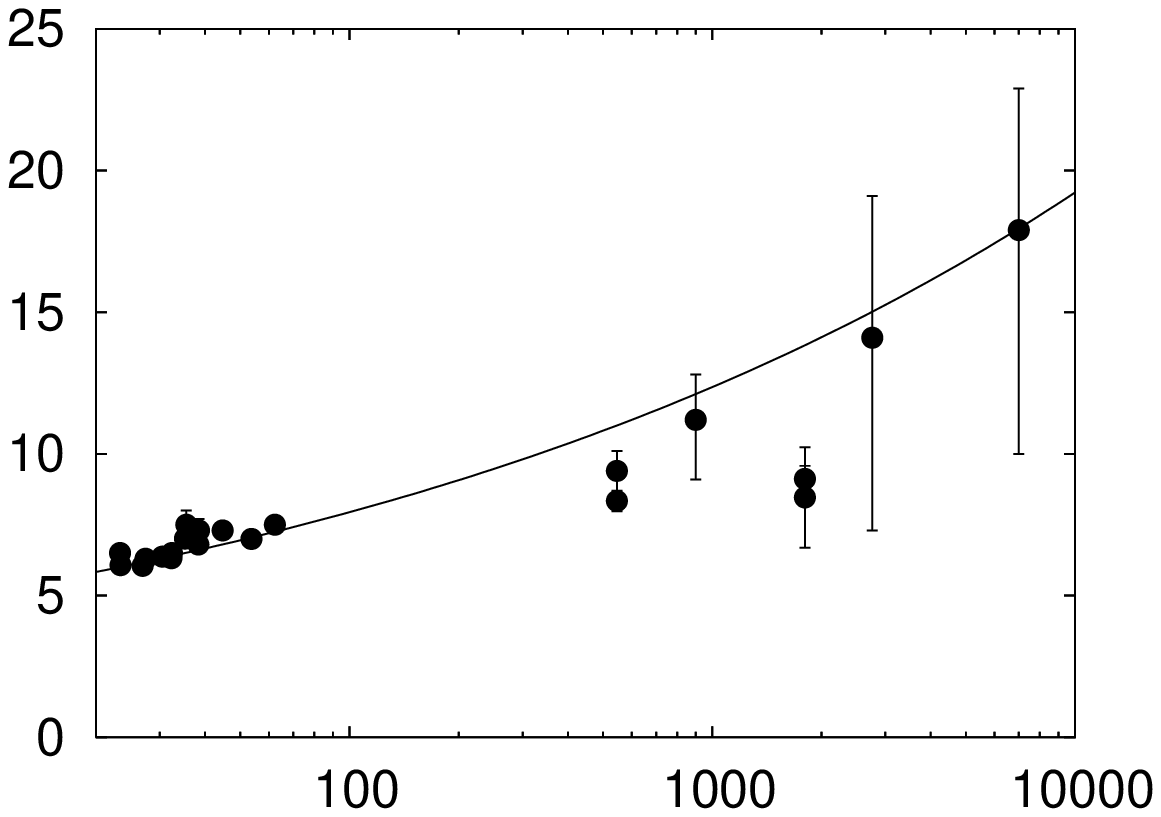}
\begin{picture}(0,0)
\setlength{\unitlength}{1mm}
\put(-8,42){$\sigma^{\rm Diff}\\  (mb)$}
\put(40,0){$\sqrt s (\rm{GeV})$}
\end{picture}
\end{minipage}
\caption{Integrated diffractive dissociation cross section for $\xi \leq 0.05$
in $p p$ and $\bar{p} p$ interactions. The data are from fixed-target 
experiments \cite{Akimov,Cool81,Sch78}, the ISR \cite{Albrow76,Arm82}, the
$Sp\bar{p}S$ \cite{Bernard87}, the Tevatron \cite{Amos93,Abe94} and the LHC
\cite{ALICE13}. The latter have been extrapolated to $\xi=0.05$ to match 
the lower-energy data.}
\ec
\label{ddsig}
\efig

According to (\ref{diffsig}),
for a fixed lower-limit of $\xi$ the cross section (\ref{ddtot}) rises
as $(\alpha_{\P}^\prime s)^{\alpha_{\P}(0)-1}$ at sufficently large $M_X$,
while for smaller values of $M_X$ there will also be a contribution
$(\alpha_{\R}^\prime s)^{\alpha_{\R}(0)-1}$ which decreases with
increasing $s$.
But if the integration is down to fixed $M_X^2=\xi_{\rm min}s$, 
the cross section rises even more rapidly, 
because as $s$ increases the 
integration extends down to increasingly smaller values of $\xi$. Note,
however, that for (\ref{diffsig}) to be applicable $M_X$ must be larger than
a few GeV. Below that we have no theory, though at low energies we do know 
that the low-mass region is dominated by baryon resonances.

Data for the $ p p \to p X$ and $\bar{p}p \to  \bar{p} X$ cross sections are 
shown in figure \ref{ddsig} with the factor of 2 included. 
For want of better information, the curve shown behaves as
$(\alpha_{\P}^\prime s)^{\alpha_{\P}(0)-1}$. 

The different experiments
correspond to varying lower limits on $M_X$ and, as we have said, we
do not know how to correct for that.
At $\sqrt{s}=546$ GeV the integrated UA4
\cite{Bernard87} cross section for $\xi < 0.05$ is $9.4 \pm 0.7$ mb and
for $M_X > 4$ GeV and $\xi < 0.05$ it is $6.4 \pm 0.5$ mb. So the cross
section for $M_X < 4$ GeV is $3.0 \pm 0.8$ mb, that is about one third of the
fully integrated cross section at that energy. For $M_X < 3.4$ GeV TOTEM 
\cite{TOTEMlo} give a cross section of ($2.62 \pm 2.17$) mb and an upper 
limit of 6.31 mb at 95$\%$ confidence level. So the energy dependence of the
baryon-resonance contribution to single diffraction remains an open question.

\begin{table}[t]
\bc
\btab{|l|c|c|}
\hline
& cross section (mb) & mass range (GeV) \\
\hline
ALICE & 14.9${+3.4\atop-5.3}$  & $ M_X < 200$ \\
CMS   & $4.27\pm0.04{+0.65\atop-0.58}$  & $12 < M_X < 394$ \\
TOTEM &  $3.3\pm0.7$  & $8 < M_X < 350$ \\
TOTEM &  $6.5\pm1.3$ & $6.5<M_X<1100$\\
\hline
\etab
\caption{Data \cite{ALICE13,CMS12,TOTEM13} for integrated cross sections at 7~TeV over the mass ranges shown. 
The error given on the TOTEM data
\cite{TOTEM13} is a notional $20\%$.}
\label{ddsig2}
\ec
\end{table}

Data from the LHC at 7~TeV are shown in table \ref{ddsig2}.
The plot of figure \ref{ddsig} includes ALICE data \cite{ALICE13} at 
$\sqrt{s}$ = 0.9, 2.7 and 7 TeV, extrapolated by the experimentalists to 
$\xi_{\rm max}=0.05$.
The CMS data \cite{CMS12} and the preliminary TOTEM 
data \cite{TOTEM13} in the table cannot be compared with those from
ALICE because, as we have explained, the correction from adding in
the contributions from smaller values of $M_X$ cannot be calculated. 
There seems no reason yet to agree with claims \cite{Khoze14,Martin14,PP14} 
that the CMS and TOTEM data call for significant modification to pomeron 
exchange. Although, as we have explained above, these are partly taken into 
account by using an effective trajectory $\alpha_{\P}(t)$, the low-mass region 
needs to be understood properly before definite conclusions can be drawn.

\section{Central production}

There are various kinds of central production events:
\begin{itemize}
{\parskip=0pt
\item strictly exclusive production of a hadron or of a resonance 
\item completely inclusive central production of a hadron or of a resonance 
\item semi-exclusive production of a system of hadrons or resonances
\item inclusive central production with rapidity gaps}
\end{itemize}
We describe the theory for each of these types of event. They have different
energy dependences. Our discussion is restricted to soft collisions, where
the centrally-produced hadron is light and has small $p_T$. 
For a recent discussion of hard collisons, see \cite{HKRS12}.

\begin{figure}[t]
\begin{center}
\epsfxsize=0.7\textwidth\epsfbox[0 0 620 300]{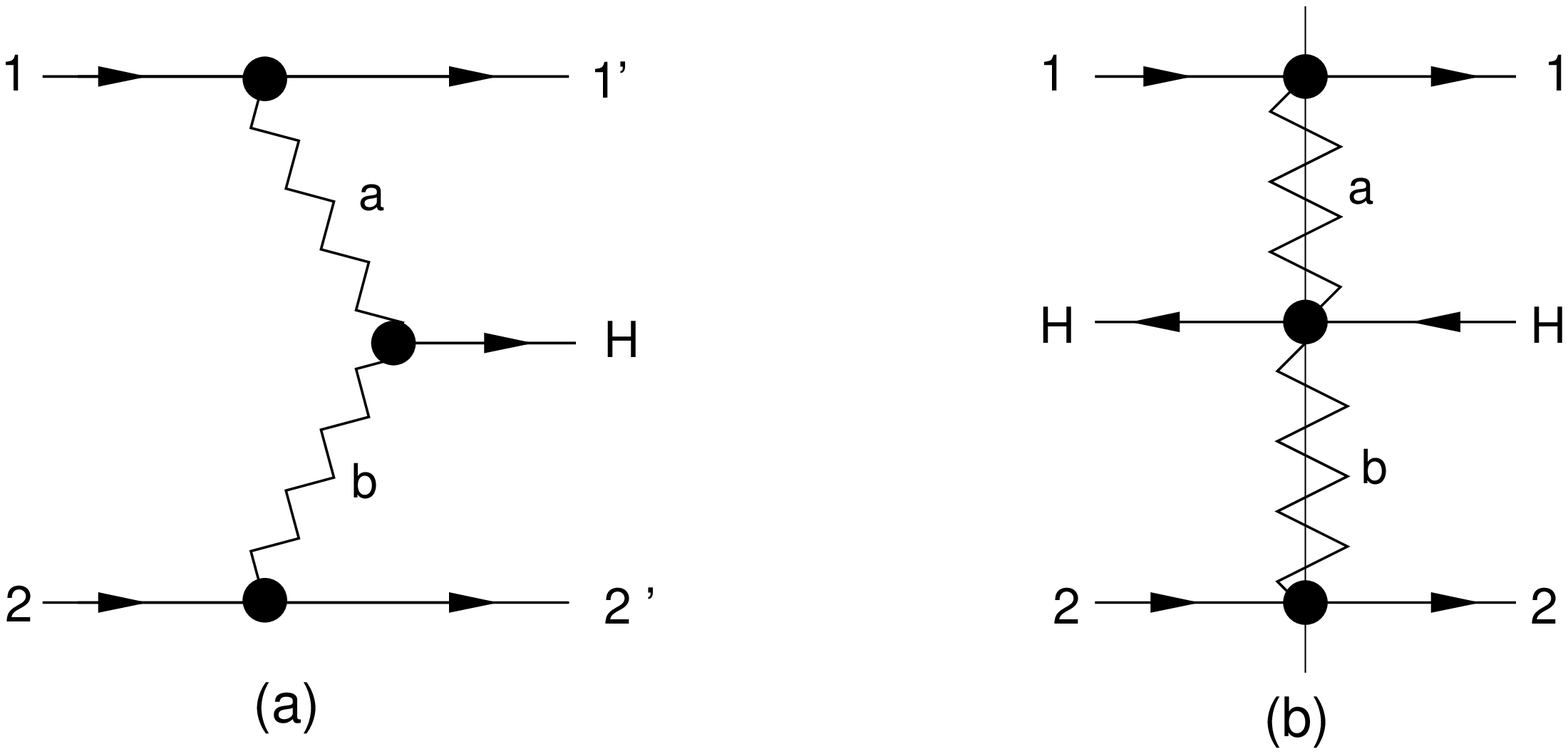}
\end{center}
\vskip -5truemm
\caption{(a) Exclusive production of a hadron or resonance $H$; (b) completely 
inclusive production}
\label{central1}
\end{figure}

\subsection{Strictly exclusive production}

For this kind of event, $AB\to A'HB'$, one calculates an amplitude, and squares
it to obtain a differential cross section. 
The amplitude depends on five independent variables, for example
\be
s=(p_1+p_2)^2 ~~~ s_1=(p_1'+p_H)^2 ~~~ s_2=(p_H+p_2')^2 ~~~ t_1=(p_1'-p_1)^2 ~~~ t_2=(p_2'-p_2)^2
\ee 
In the case of central production at high energy, both $s_1$ and $s_2$ will be large and
the amplitude is a sum of terms of 
the type shown in figure \ref{central1}a, where $H$ is the hadron or resonance being 
produced. Each of the zigzag lines $a,b$ represents any of the exchanges
$\P, \rho, \omega, f_2, a_2$ etc 
and the contribution to the amplitude from the exchanges of reggeons $a,b$ is
\beqa
&&\beta_{Aa}(t_1)~\xi_{a}(t_1)
~\beta_{Bb}(t_2)~\xi_{b}(t_2)
~f_{abH}(\eta, t_1,t_2)
~(\alpha'_as_1)^{\alpha_a(t_1)}
(\alpha'_bs_2)^{\alpha_b(t_2)}.
\label{excl}
\eeqa
Here $\eta=s_1s_2/(sM_H^2)$. The functions $\beta(t)$ are the same as occur in 
the contribution (\ref{elastic}) to the elastic scattering amplitude, but this 
information  is not very useful because not a great deal is known about the 
(complex) function $f_{abH}(\eta, t_1, t_2)$ that couples the reggeons to the 
hadron $H$ -- its dependence on $t_1$ and $t_2$ is completely unknown and
its $\eta$ dependence is known \cite{DLZ69} only for small $\eta$. 

In most high-energy events  $t_1$ and 
$t_2$ will be small. Then, with  $\xi_1,\xi_2$ the fractional longitudinal 
losses of the initial hadrons, 
\be
s_1\sim s~\xi_2  ~~ ~~~~ s_2\sim s~\xi_1 ~~ ~~~~ \xi_1\xi_2s\sim M_H^2 ~~ ~~~~ \eta\sim 1
\ee
Hence dependence on one of the variables, say the angle between the final-state particles $A'$ and $B'$ has disappeared, and the energy dependence of the amplitude (\ref{excl}) is given by the factor
\be
(\alpha'_a\xi_2s)^{\alpha_a(t_1)}
(\alpha'_b\xi_1s)^{\alpha_b(t_2)}=
(\alpha'_a\xi_2s)^{\alpha_a(t_1)}
(\alpha'_bM_H^2/\xi_2)^{\alpha_b(t_2)}
\label{endep1}
\ee
To obtain the energy dependence of the cross section, one has to square this and
apply
\be
\int_{M_H^2/(s \xi_{max})}^{\xi_{max}} d\xi_2/\xi_2.
\ee
The result is 
\beqa
&&\frac{1}{2(\alpha_a(t_1)-\alpha_b(t_2))}
\Big\{(\alpha'_a s)^{2\alpha_a(t_1)}
(2\alpha'_bM_H^2)^{\alpha_b(t_2)}(\xi_{max})^{2(\alpha_a(t_1)-\alpha_a(t_2))}
\nonumber\\
&&-(\alpha'_b s)^{2\alpha_b(t_2)}
(\alpha'_aM_H^2)^{2\alpha_a(t_1)}(\xi_{max})^{2(\alpha_a(t_2)-\alpha_b(t_1))}\Big\}
\label{endep2}
\eeqa

So, up to logarithmic factors, the $\P\P$, $\P\R$
and $\R\P$ term give a
contribution increasing as $s^{2\epsilon_{\P}}$, with $\epsilon_{\P}$ given in
(\ref{pomtrajectory}),
while for the $R\R$
term the power is  close
to -1. However, when we square the amplitude there are also interference terms
\cite{Kirk00}, including one behaving approximately as $1/\sqrt{s}$.

The principal reason for studying central production of mesons in high energy
$p p$ collisions is to search for glueballs, mesons consisting of two or three
 ``constituent'' gluons. Reactions of the type $p p \to p X^0 p$, 
in which the exchanges $a$ and $b$ of figure \ref{central1}
are both pomerons, are thought most likely to produce glueballs,  as the pomeron is believed to be primarily a gluonic system 
\cite{Meyer05}. 

States with two constituent gluons necessarily have $C=+1$. The lowest-mass
states are expected to be in a relative $S$-wave and have the quantum numbers 
$J^{PC} = 0^{++}$, $2^{++}$ and $0^{-+}$. Three-gluon systems can have both 
$C=+1$ and $C=-1$ and for the ground states with a relative $S$-wave the 
quantum numbers are $J^{PC} = 1^{++}$, $1^{+-}$, $1^{--}$ and $3^{--}$.
Conventional $q\bar{q}$ states exist in all of these $J^{PC}$ states which
presents a problem as there is naturally mixing between the two systems. The
presence of glueballs has to be inferred by an excess of states with 
specific glueball quantum numbers.   

Masses of glueballs have been estimated in quenched lattice gauge theory
\cite{Bali93,MP99,Chen06} and the results are given in table \ref{LQCD}.

\begin{table}[t]
\bc
\btab{|l|l|l|l|}
\hline
$J^{PC}$ & Bali {\it et al} \cite{Bali93} & Morningstar {\it et al} \cite{MP99}
& Chen {\it et al} \cite{Chen06} \\
\hline 
$0^{++}$ & $1550 \pm ~30$ & $1730 \pm 50$ & $1709 \pm 49$ \\
$2^{++}$ & $2270 \pm 100$ & $2400 \pm 25$ & $2388 \pm 23$ \\
$0^{-+}$ & $2330 \pm 260$ & $2590 \pm 40$ & $2557 \pm 25$ \\
\hline
\etab
\ec
\vskip -1pt
\caption {Glueball masses in MeV in quenched-lattice approximation.}
\label{LQCD}
\end{table}

The $2^{++}$ and $0^{-+}$ are at the limit of current meson spectroscopy, so 
experimental and theoretical emphasis has been on the scalar mesons $f_0(980)$,
$f_0(1370)$, $f_0(1500)$ and $f_0(1710)$. In terms of the standard $q\bar{q}$
model assignments for the light mesons there is one too many isoscalar scalars
in the 1300 to 1700 MeV mass region and this has been attributed to mixing with
a scalar glueball. However it has been argued that the $f_0(1370)$ may not
exist, although this has been strongly contested and the situation remains 
unclear \cite{KZ07}. Removing this ambiguity and extending meson spectroscopy 
to higher mass are key to unravelling the glueball question. 

Preliminary data from ALICE \cite{Schicker12} on $\pi^+\pi^-$ production in 
$p p$ collisions at $\sqrt{s} = 7$ TeV with a large double gap in 
pseudorapidity show dominant peaks in the $\pi^+\pi^-$ mass distribution 
associated with the $f_0(980)$ and $f_2(1270)$. While it is 
tempting to conclude that this result illustrates double-pomeron exchange 
preferentially selecting isoscalar states, without knowing the detailed 
kinematics this is somewhat premature. The $f_2(1270)$ is a well-establishd 
$q\bar{q}$ state and the $f_0(980)$ is not generally considered as a gluonic
system. Further there is no evidence for the $f_0(1500)$ which is the scalar
meson thought most likely to have a large gluonic component. Given the 
discussion after (\ref{endep2}) it may be that the $\R\P+\P\R$ terms are 
responsible.  

A detailed review of the present status of glueballs can be found in 
\cite{Ochs13}.

A second reason for studying exclusive meson production is to continue the 
search for evidence of the elusive odderon,
the possible $C = P = -1$ partner of the $C = P = +1$ pomeron. The odderon has
a long history \cite{Ewerz03}, but unambiguous odderon effects have 
never been observed. It has been suggested \cite{Otto91} that the odderon  
could be observed in high-energy $pp$ interactions through exclusive central 
production of $C = -1$ mesons, for example the $J/\psi$. 
The proposed mechanism is figure \ref{central1} with reggeon $a$ the pomeron 
and reggeon $b$ the odderon.
Note, however, that if we take $b$ to be the $\omega$ trajectory, or even
the $J/\psi$, the energy dependence would be the same. Which of the choices
dominates is determined by the relative magnitudes of the couplings to the 
$J/\psi$ and the lower proton.

As an
alternative to an exclusive process, the difference in inclusive production
of particles and antiparticles in the central region has been proposed
\cite{Merino09} and present data analysed, without success. Reasons for the 
missing odderon are given in \cite{DDN06}, including its weak coupling to the 
nucleon at small $t$. A further complication arises in lattice gauge 
theory \cite{Meyer05} as it predicts that the odderon trajectory, although 
it has a slope similar to that of the pomeron, $\epsilon_{\rm Odd}$ is negative
so the odderon has an intercept that is less than one.

\subsection{Completely inclusive production}

In this case, $AB\to HX$, the diagram is figure 3b. Each reggeon carries zero 
4-momentum. The vertical line indicates that, 
according to the generalised optical theorem \cite{Mueller70}, to
calculate the inclusive differential cross section one has to take the
discontinuity in the variable $(p_1+p_2-p_H)^2$.

Define the momentum transfers 
\be
\tau_1=(p_1-p_H)^2 ~~~~ \tau_2=(p_2-p_H)^2
\ee
and let the fraction of the initial centre-of-mass-frame momentum  of hadron 
$p_1$ carried by $H$ be $x$. Then figure 3b contributes
\be
{d^2\sigma\over {d\log x~dP_{TH}^2}}={\beta_{Aa}(0)\beta_{Bb}(0)V_{abH}\over 
{4\pi^3}s}
~\big (-\alpha_a'\tau_1\big)^{\alpha_a(0)}\big (-\alpha_b'\tau_2\big)^
{\alpha_b(0)}
\label{inclusive}
\ee
where we have used $d^3P_H/E_H\sim \pi~\!\!dx~\!\!dP^2_{TH}/x$.
The coupling $V_{abH}$ of the two reggeons to the hadron $H$ is a constant
since, as the
reggeons carry zero 4-momentum, the only Lorentz invariant at the central 
vertex is $P_H^2$.

If the exchanges $a$ and $b$ are the same, and also the initial hadrons have 
the same mass $m$,
\be
\big(-\alpha'\tau_1\big)^{\alpha(0)}\big(-\alpha'\tau_2\big)^{\alpha(0)}\sim
\big(\alpha's\big)^{\alpha(0)}\big(\alpha'(m_H^2+p_{TH}^2+x^2m^2)\big )^
{\alpha(0)}
\ee

\bfig[t]
\bc
\epsfxsize=0.4\textwidth\epsfbox[0 0 440 400]{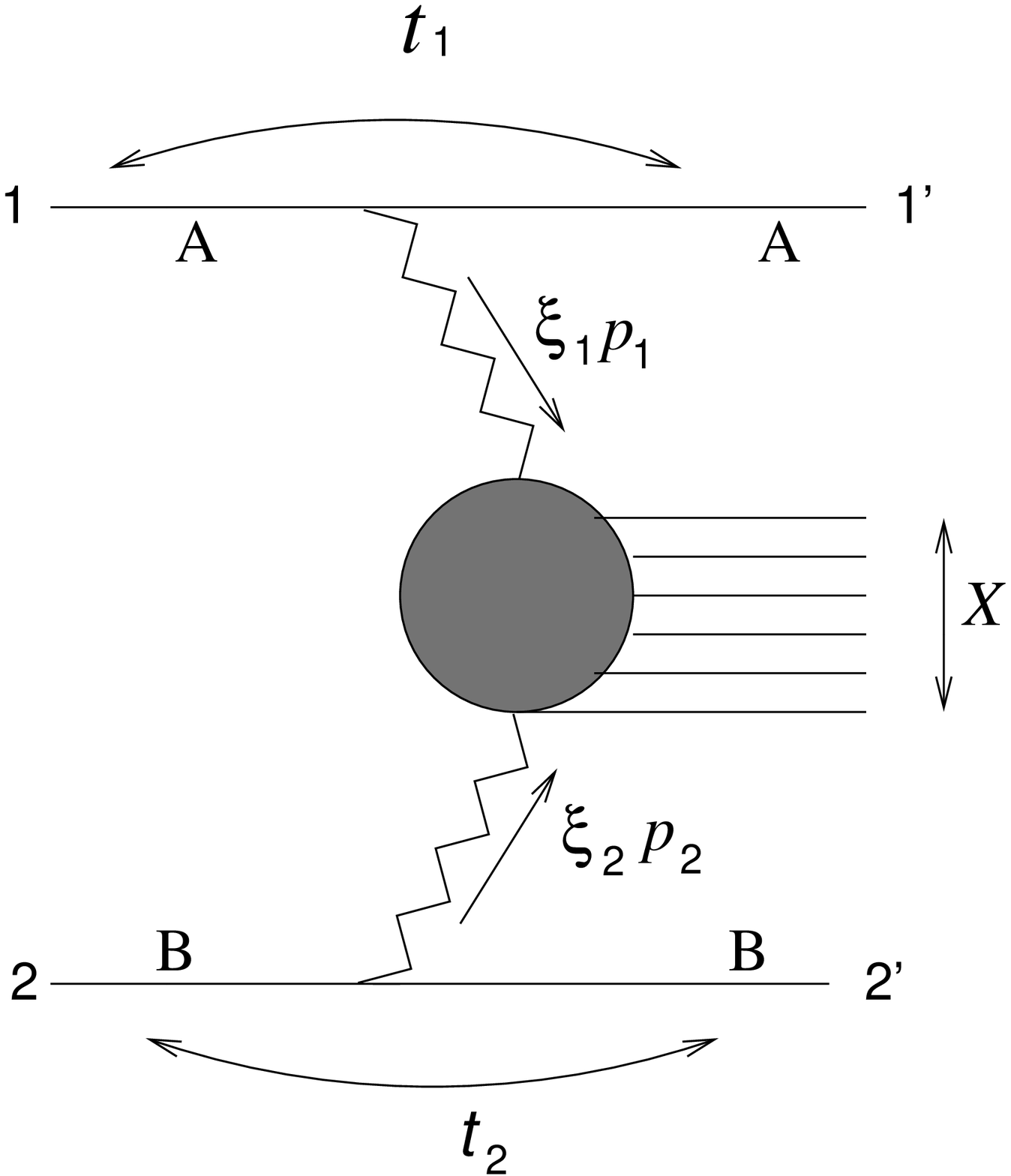}
$\phantom{XXXX}$
\epsfxsize=0.35\textwidth\epsfbox[0 0 350 400]{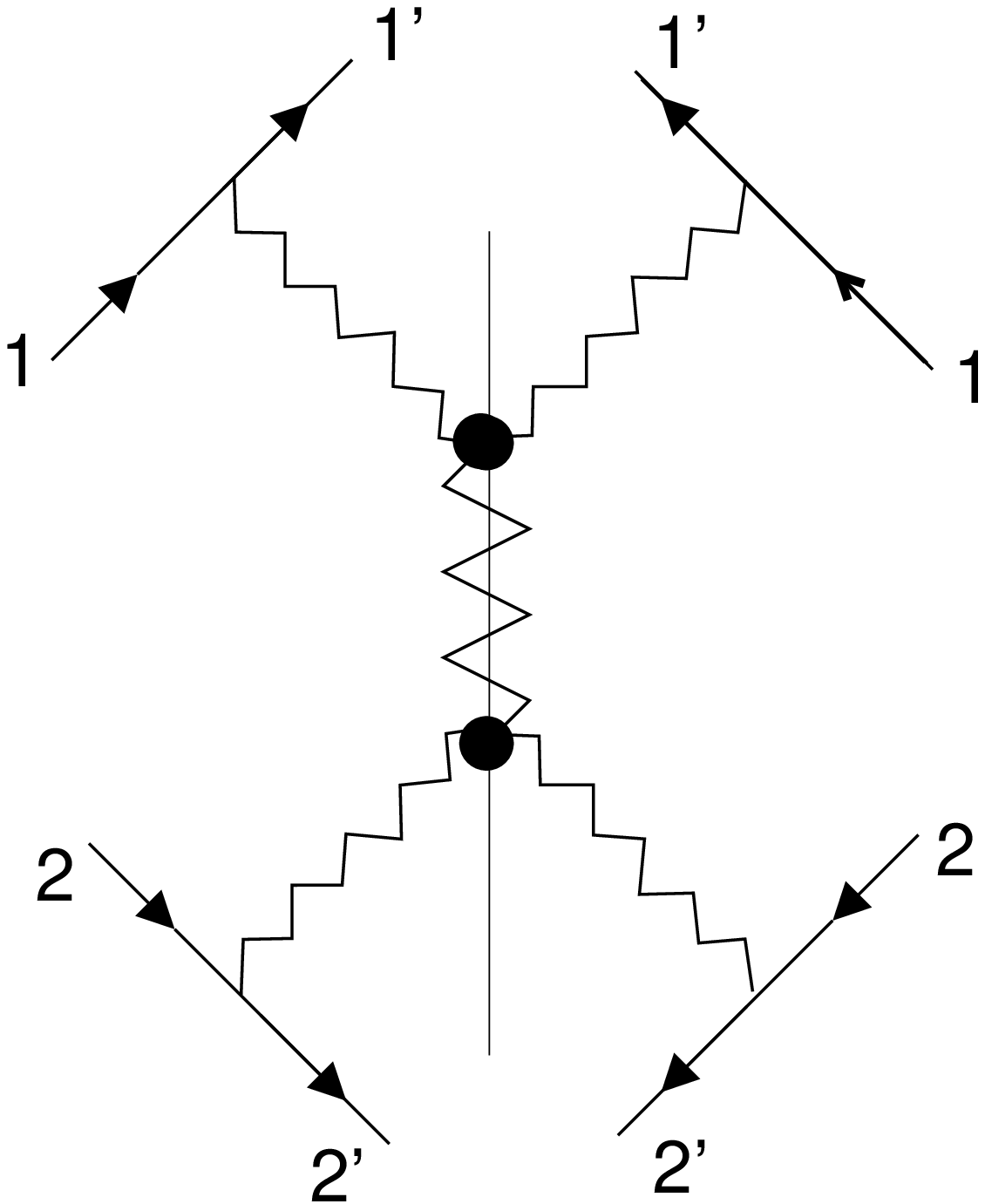}
$\hbox{(a)}\phantom{XXXXXXXXXXXXXXXXX}\hbox{(b)}$
\caption{(a) The inclusive process (\ref{inclusive1}) for the case where
both initial particles lose very little momentum. (b) The squared amplitude
of (a) summed over $X$ when the invariant mass of the system $X$ is large --
the vertical line indicates that the discontinuity must be taken in the 
variable $M_X^2$.}
\label{twopom}
\ec
\efig

\bfig[t]
\bc
\epsfxsize60mm
\epsffile{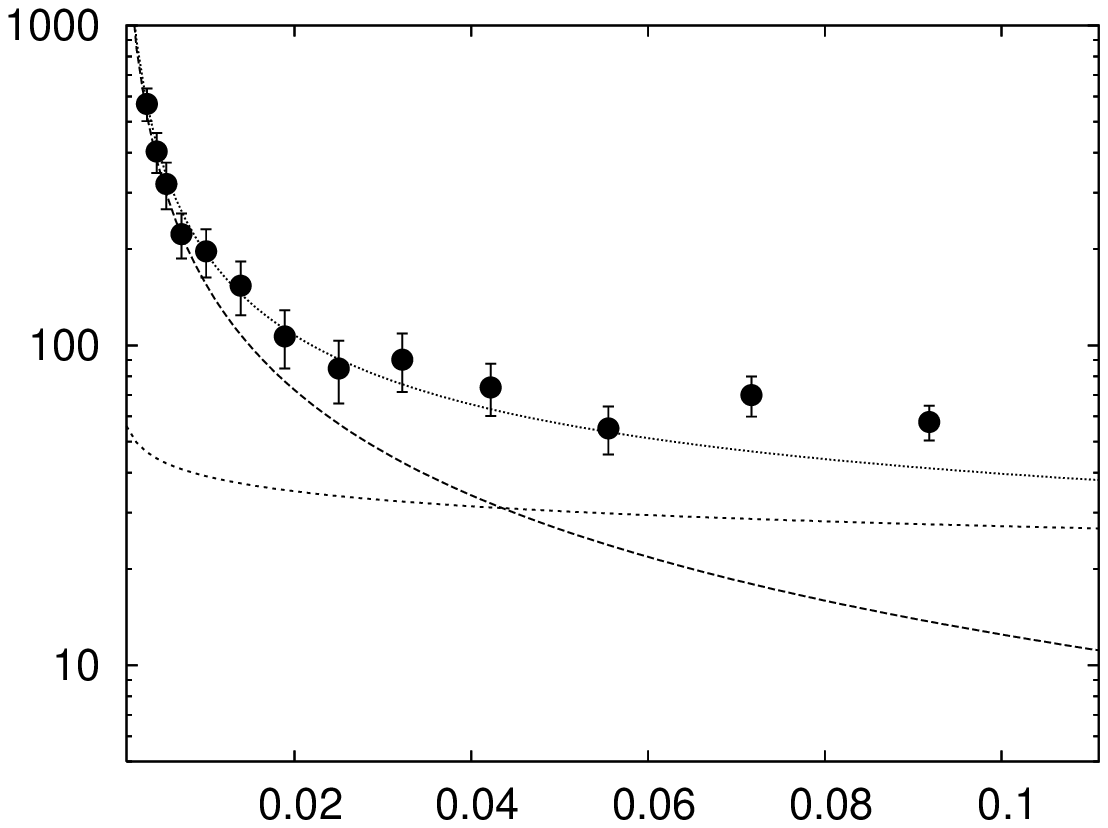}
\epsfxsize60mm
\epsffile{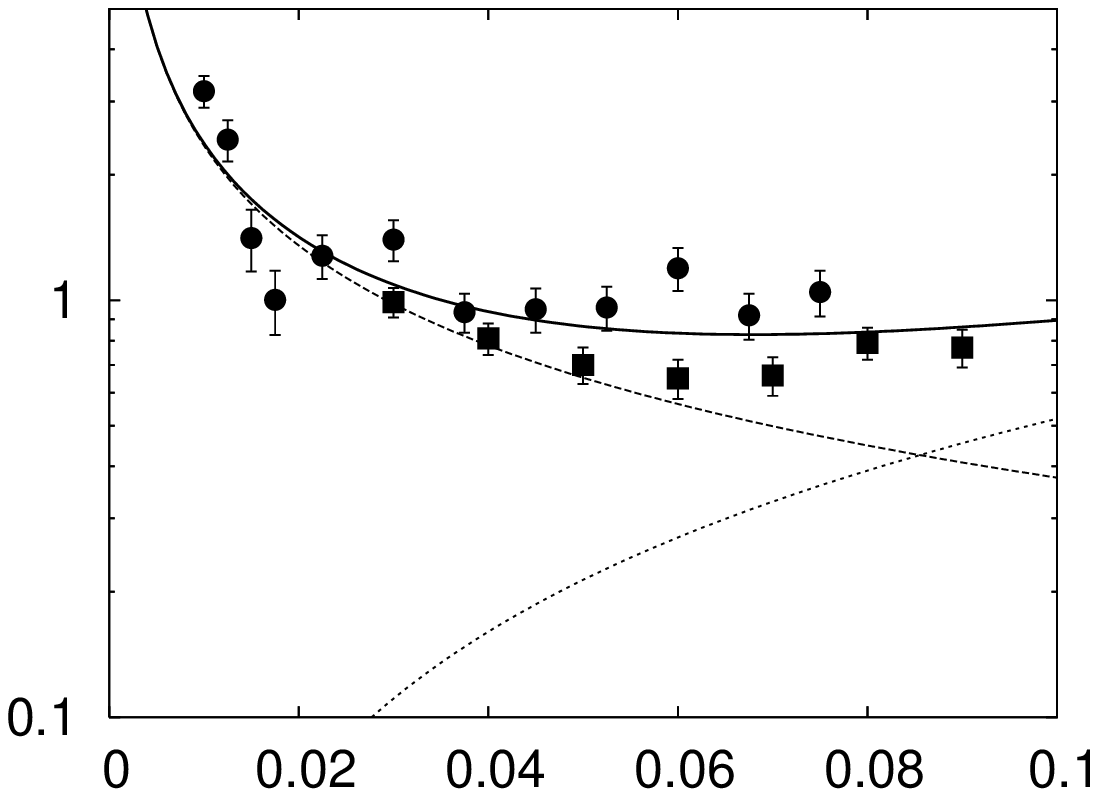}
\vskip -5pt
$~~~~~~~~~~~~~~~~~~~~\xi~~~~~~~~~~~~~~~~~~~~~~~~~~~~~~~~~~~~~~~~~~~~~~~~\xi$
\vskip 1 pt
$\phantom{XXX}$(a)$\phantom{XXXXXXXXXXXXXXXXXX}$(b)
\caption{The double differential cross section $d^2\sigma/dtd\xi$
in $\bar{p} p$ interactions at (a) $\sqrt{s} =$ 1800 GeV, $t = -0.05$ GeV$^2$
and (b) $\sqrt{s} =$ 540 GeV (circles) and 630 GeV (squares), $t = -0.95$
GeV$^2$  The data are from \cite{GM99,Bozzo84,Brandt98}, the dashed line is
the $\P\P\P$ contribution, the dotted line the $\R\R\P$ contribution and the
solid line their sum. The units are mb~GeV$^{-2}$.}
\label{cdf}
\vskip 18pt
\epsfxsize59mm
\epsfbox[50 50 390 290]{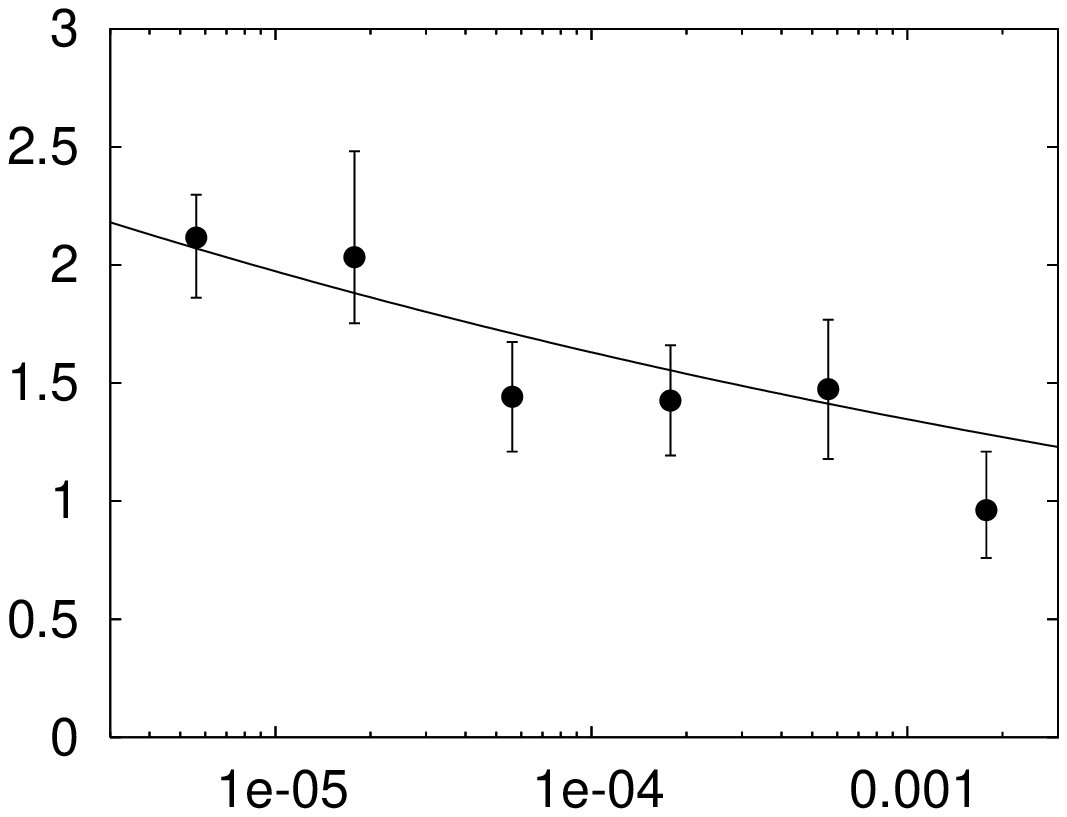}
\epsfxsize59mm
\epsfbox[50 50 390 290]{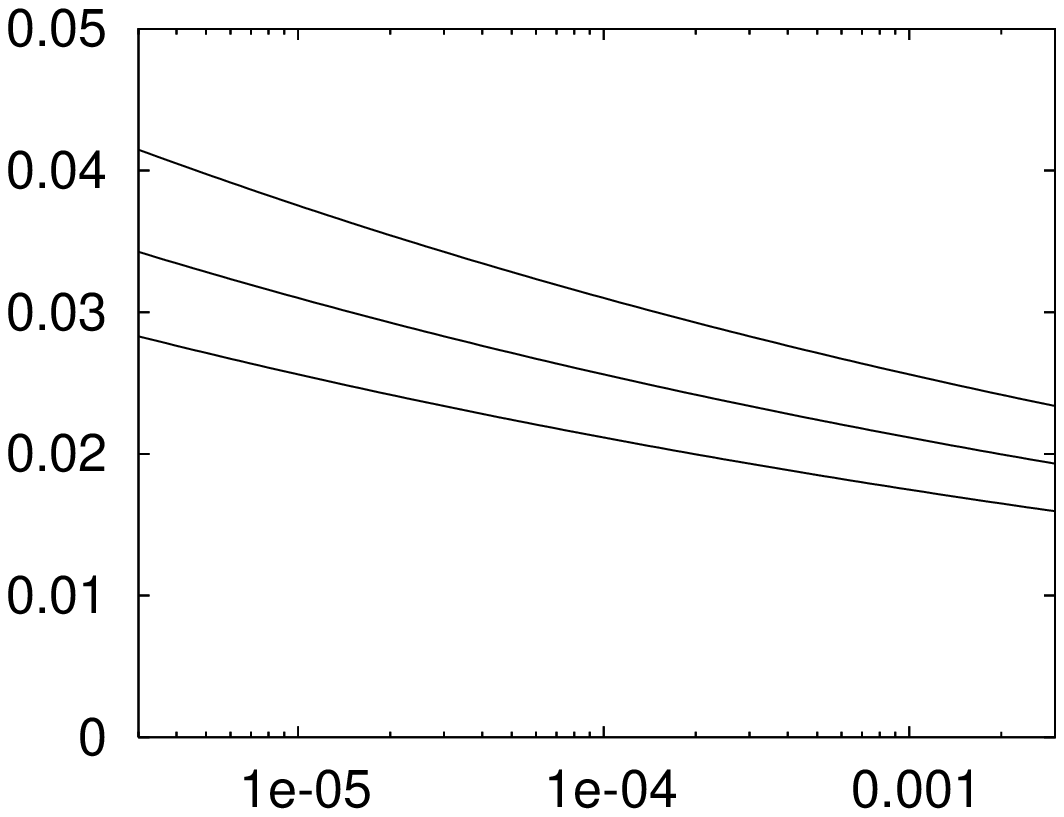}
\vskip -5pt
$~~~~~~~~~~~~~~~~~~~~\xi~~~~~~~~~~~~~~~~~~~~~~~~~~~~~~~~~~~~~~~~~~~~~~~~\xi_1$
\vskip 1 pt
$\phantom{XXX}$(a)$\phantom{XXXXXXXXXXXXXXXXXX}$(b)
\caption{(a) Data \cite{CMS12} for $\xi d\sigma/d\xi$ at 
$\sqrt{s} = 7$ TeV; the units are mb.
(b) $\xi_1\xi_2d^2\sigma/d\xi_1~d\xi_2$ at $\sqrt{s} = 7$ TeV, for 
$\xi_2 = 10^{-5}$ (top curve), $10^{-4}$ (middle curve) and $10^{-3}$ 
(bottom curve); the units are mb.
}
\label{cmsd2}
\ec
\efig
\subsection{Semi-exclusive production}

Figure \ref{twopom}a shows the particle $H$ in figure 3(a)  replaced with a cluster of particles to
give the process
\be 
A(p_1)+B(p_2) \to A(p_1^\prime)+B(p_2^\prime)+X
\label{inclusive1}
\ee  
For most events both $t_1$ and $t_2$ will be small and
if both initial particles lose a small amount of energy 
$p_1^\prime
\sim (1-\xi_1) p_1$ and  $p_2^\prime \sim (1-\xi_2) p_2$ with 
$\xi_1,\xi_2  \ll 1$.  If $\xi_1$ and 
$\xi_2$ are sufficiently small then both reggeons will be 
pomerons. If we square the amplitude and sum over all possible systems $X$ then
\be
\frac{d^4\sigma}{dt_1d\xi_1dt_2d\xi_2}=D^{\P/A}(t_1,\xi_1)D^{\P/B}(t_2,\xi_2)\sigma^{\P\P}
(M_X^2,t_1,t_2)
\label{d4sig_1}
\ee
where on the right-hand side the first two terms are the pomeron-flux factors of
(\ref{diffdissoc}), and $\sigma^{\P\P}(M_X^2)$ is the total cross section for pomeron-pomeron 
scattering with $M_X^2 = \xi_1\xi_2s$. If $M_X^2$ is sufficiently large then 
only pomeron exchange between the pomerons need be included so that the 
right-hand side of (\ref{d4sig_1}) corresponds to figure \ref{twopom}b. 
This contains two triple-pomeron vertices and the diagram factorises, allowing 
the cross section (\ref{d4sig_1}) to be written as
\be
\frac{d^4\sigma(s)}{dt_1d\xi_1dt_2d\xi_2}=\frac{d^2\sigma(s)}{dt_1d\xi_1}
\frac{d^2\sigma(s)}{dt_2d\xi_2}\frac{1}{\sigma^{\rm Tot}_{\P}(s)}
\label{d4sig_2}
\ee
Here $\sigma^{\rm Tot}_{\P}(s)$ is the pomeron-exchange contribution to
the $pp$ total cross section.

This formula still applies when either or both of the upper reggeons in
figure \ref{ddpom}b for the 
first two factors on the right-hand side represent non-pomeron exchanges. Indeed, unless $\xi$ is extremely 
small two terms are required at high energy to describe the double distribution
 $d^2\sigma/dtd\xi$. These are $\P\P\P$ and $\R\R\P$. (As we have said before,
the interference terms $\R\P\P$ and $\P\R\P$ are usually omitted for simplicity.) 
At fixed $\xi$, both have the same $s$ 
dependence as can be seen from (13); however they have very different 
$\xi$ dependence. At small $t$, $\P\P\P$ behaves approximately as $1/\xi$ and 
$\R\R\P$ behaves approximately as a constant. As $t$ increases the $\xi$ 
dependence of the $\P\P\P$ term changes slowly because of the small slope of
the pomeron trajectory. However the much larger slope of the nonleading
trajectory $\R$ causes a rapid change in the $\xi$ dependence of the $\R\R\P$ 
term so that it behaves approximately as $\xi^{1.3}$ at $t = -1$ GeV$^2$.

We illustrate
this explicitly with the CDF data \cite{GM99} for $d^2\sigma/dtd\xi$  at 
$\sqrt{s} =$ 1800 GeV, $t = -0.05$ GeV$^2$ and the UA4 and UA8 data at 
$t=-0.95$ GeV$^2$ and $\sqrt{s} =$ 546 and 630 GeV respectively
\cite{Bozzo84,Brandt98}. For the $\xi$ dependence we use the pomeron and 
$C=+$ reggeon parameters of (\ref{parameters}),
though different choices give qualitatively the same outputs.
The result is shown in figure \ref{cdf}, the normalisation of the two 
terms in each case being adjusted to give a reasonable description of the 
data. At $t = -0.05$ GeV$^2$ pomeron dominance of $d^2\sigma/dtd\xi$ only 
occurs for $\xi \lesssim 0.004$. However at $t=-0.95$ GeV$^2$ it occurs for 
$\xi\lesssim 0.03$. 

An estimate of $d^4\sigma/dt_1d\xi_1dt_2d\xi_2$ at LHC energies
can be made by combining the single-diffraction TOTEM data \cite{TOTEM13} for 
$d^2\sigma/dt$ for $8 < M_X < 350$ with the CMS data \cite{CMS12} for 
$d^2\sigma/dtd\xi$ over essentially the same mass range: see table 
\ref{ddsig2}. The corresponding values of $\xi$ are sufficiently small, 
$\xi \lesssim 0.003$,  to ensure triple-pomeron dominance of the 
single-diffractive cross section. The exponential slope of the TOTEM data over 
this mass range is 8.5 GeV$^2$, giving a mean value of $t = -0.082$ GeV$^2$ 
which can be used to calculate the shape of $d^2\sigma/dtd\xi$. The result is 
shown in figure \ref{cmsd2}a, with the 
normalisation adjusted to fit the data. The result of applying 
(\ref{d4sig_2}) is given in figure \ref{cmsd2}b.
\subsection{Inclusive central production with rapidity gaps}

\begin{figure}[t]
\begin{center}
\epsfxsize=0.75\textwidth\epsfbox[0 0 705 565]{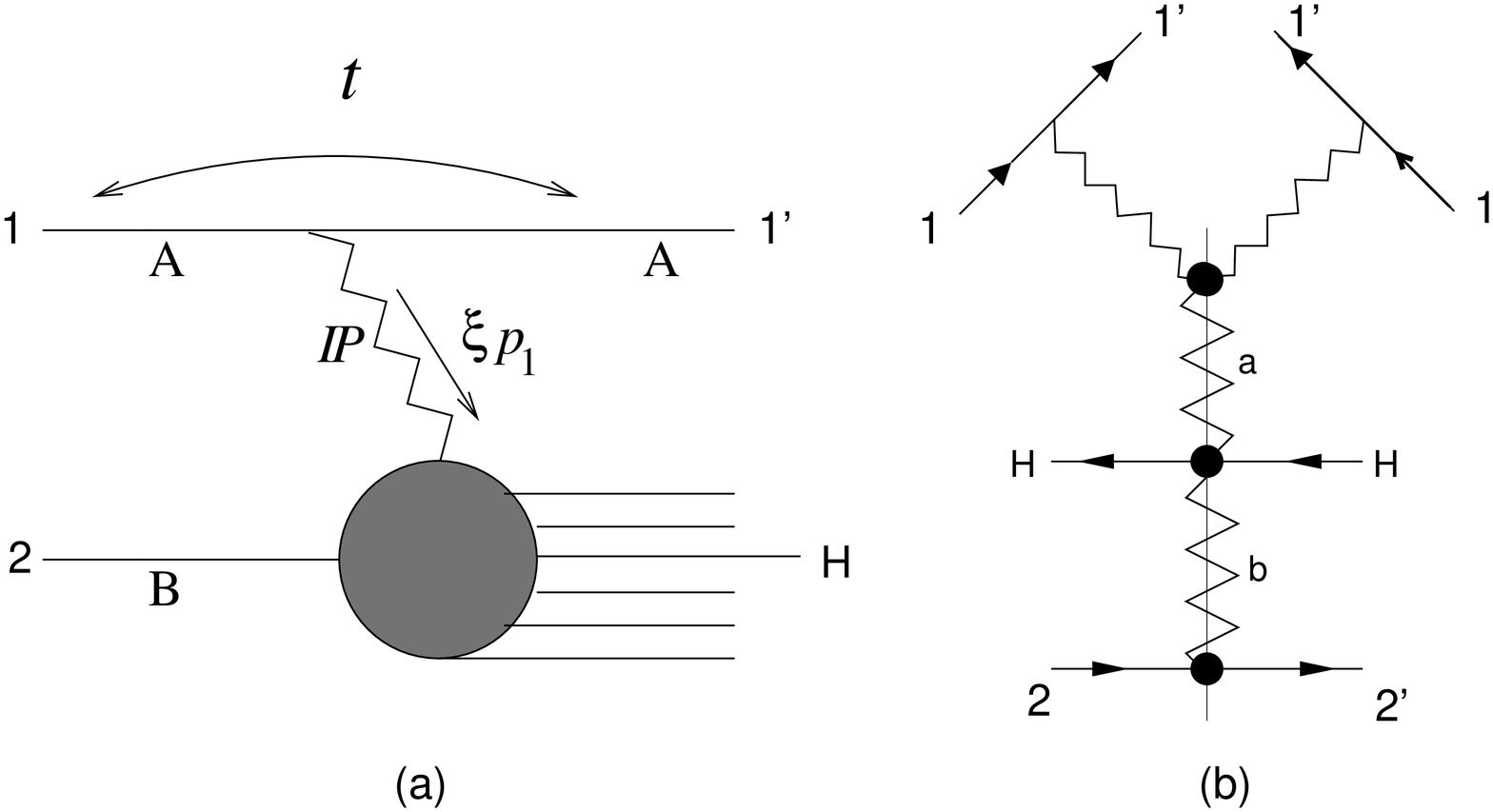}
\end{center}
\caption{(a) Inclusive central production  of a hadron $H$ with one initial hadron
losing very little momentum; (b) the square of the amplitude summed over
possible systems of particles accompanying $H$.}
\label{rapgap}
\end{figure}

If the initial hadron $p_1$ loses only a small fraction $\xi_1$ of its initial
momentum the mechanism is that of figure \ref{rapgap}a. 
If $\xi_1$ is extremely small, energy conservation will demand that there
be a rapidity gap between $p_1'$ and the rest of the final-state particles.
Whether or not that is the case, if $\xi_1$ is small 
enough and the total energy is high enough, the dominant contribution will 
come from all the reggeons being pomerons and factorisation will apply:
\be
{d^4\sigma(s)\over {dt_1d\xi_1d\log x~dP_{TH}^2}}=\frac{d^2
\sigma(s)}{dt_1d\xi_1}~{d^2\sigma(s)\over {d\log x~dP_{TH}^2}}\frac{1}{\sigma^{\rm Tot}(s)}
\ee
where the first factor on the right-hand side is that of (\ref{diffdissoc}) 
and the second that of (\ref{inclusive}). 
 
Similarly, if both initial hadrons lose only a very small fraction of their
initial momenta, 
\be
{d^6\sigma(s)\over {dt_1d\xi_1dt_2d\xi_2d\log
x~dP_{TH}^2}}=\frac{d^4\sigma(s)}{dt_1d\xi_1dt_2d\xi_2}~{d^2\sigma(s)\over {d\log
x~dP_{TH}^2}}\frac{1}{\sigma^{\rm Tot}(s)}
\ee
where the first factor on the right-hand side is that of (\ref{d4sig_2}).

\end{document}